\documentclass[aps,preprint,superscriptaddress,showpacs,floatfix]{revtex4}
\usepackage{amssymb}
\usepackage[bf]{caption}
\usepackage{graphicx}
\usepackage{amsmath}
\usepackage{wrapfig}

\begin{document}

\title{Isoscaling and the high Temperature limit}
\author{C. O. Dorso}
\affiliation{Universidad de Buenos Aires, Nu\~nez, Argentina}
\author{C. M. Hern\'andez}
\affiliation{Universidad de Colima, Colima, M\'exico}
\author{J. A. L\'opez}
\affiliation{The University of Texas at El Paso, El Paso, Texas 79968, USA}
\author{J. A. Mu\~noz}
\affiliation{The University of Texas at El Paso, El Paso, Texas 79968, USA}

\date{\today}

\begin{abstract}
This study shows that isoscaling, usually studied in nuclear reactions, is a
phenomenon common to all cases of fair sampling. Exact expressions for the
yield ratio $R_{21}$ and approximate expressions for the isoscaling
parameters $\alpha$ and $\beta$ are obtained and compared to experimental
results. It is concluded that nuclear isoscaling is bound to contain a
component due to sampling and, thus, a words of caution is issued to those
interested in extracting information about the nuclear equation of state
from isoscaling.
\end{abstract}

\maketitle

\section{Introduction}

\label{intro} Recent studies of the isospin dependence of nuclear reactions
at intermediate energies have been studied by comparing fragmenting
collisions of similar mass and energies but different isospin.
Experimentally, the ratio of isotope yields in two different systems, $1$
and $2$, $R_{21}(n,z)=Y_{2}(n,z)/Y_{1}(n,z)$, has been seen to follow,
approximately, an exponential function of the neutron number, $n$, and the
proton number, $z$, of the isotopes; dependence known as \textit{isoscaling}~%
\cite{tsang2001,tsang2001a,Raduta2005}: 
\begin{equation}
R_{21}=Y_{2}(n,z)/Y_{1}(n,z)=C\exp (\alpha n+\beta z),  \label{eq1}
\end{equation}%
where $\alpha $ and $\beta $ are fitting parameters and $C$ is a
normalization constant. This fit has been studied under the light of several
theoretical models of nuclear reactions and, under different assumptions
(see \textit{eg.}~\cite{botvina}), the fitting parameters can be expected to
be related to the symmetry term of the equation of state, $C_{sym}$, through 
\begin{equation}
\alpha =\frac{4C_{sym}}{T}\left[ \left( Z_{1}/A_{1}\right) ^{2}-\left(
Z_{2}/A_{2}\right) ^{2}\right] \ ,  \label{eq2}
\end{equation}%
\ where $T$ is the assumed temperature of both reactions Thus the interest
on studying the isoscaling phenomenon: $R_{21}$ has the potential of
elucidating the behavior of $C_{sym}$ at varying isospin, Temperature, etc..
It is relevant to notice that according to the above displayed relation the
coefficient $\alpha $ should approach $0$ as $T$ increases.

In a series of recent works, however, it has been shown that the phenomenon
of isoscaling can be found very early --before thermalization-- in classical
molecular dynamics simulations of nuclear reactions~\cite{dorso-CMD}, as
well as in non-thermal physical phenomena, such as in percolating networks.
In particular, percolation in two \textquotedblleft \textit{colors}%
\textquotedblright , (\textit{i.e.} protons and neutrons~\cite{davila}), or
in extended \textquotedblleft polychromatic\textquotedblright\ nets~\cite%
{dorso-1}, has demonstrated that the isoscaling behavior of the form of
equation~(\ref{eq1}) emerges as a direct consequence of simple combinatorial
problems with, \textit{eg.} $\alpha =\ln (q_{2}/q_{1})$, where $%
q_{i}=N_{i}/A_{i}$. These results point to the fact that isoscaling,
although connected to the equation of state, can also be produced by
non-thermal processes.and then, probabilistic aspects of the problem can
play a disturbing role complicating the determination of $C_{sym}$ from
isoscaling deermination. Thus the motive of the present work: to study the
probabilistic aspects of isoscaling.

In this study we investigate the phenomenon of isoscaling in the very
simplest scenario --free of the geometrical constraints imposed by bond
percolation-- of sampling \textquotedblleft protons\textquotedblright\ and
\textquotedblleft neutrons\textquotedblright\ directly from an urn. In
section~\ref{sec1} we will obtain the correct isoscaling law using
probabilistic arguments, and will show that equation~(\ref{eq1}) is a
limiting approximation to the exact expression. In section~\ref{sec2} we
compare our results to experimental values, after which the manuscript
closes listing several conclusions.

\section{Isoscaling and sampling}

\label{sec1}

Consider the problem of building clusters containing a number $a$ of
nucleons by simply grabbing these $a$ particles from an urn in which there
are $A$ particles composed of $Z$ ``protons'' and $N$ ``neutrons'', $i.e. \
A=N+Z$. We assume the sampling to be without replacement, and with no
interactions amongst the particles nor with the urn itself; these premises
are known as simple random sampling ($SRS$) in the statistics circles.

To use this setup to study isoscaling we first focus on determining the
yield of fragments, $Y(n,z)$, that the previous scheme would produce after a
large number of samplings. Stands to reason that such yield would be
directly related to the probability of drawing $n$ neutrons and $z$ protons, 
\textit{i.e.} $a=n+z$. Repeating then for a second urn with a different
isotopic composition, one can easily obtain the corresponding $R_{21}$ and,
thus, the scaling law.

Let us first determine the probability of ending with a cluster composed by $%
n$ neutrons and $z$ protons, \textit{i.e.} $a=n+z$. The number of ways in
which a cluster of $a$ particles can be obtained from randomly sampling an
urn with $A$ particles is $\binom{A}{a}=A!/(a!(A-a)!)$. Out of these
possibilities only $\binom{N}{n}\times \binom{Z}{z}$ will correspond to
clusters with $n$ neutrons and $z$ protons. Thus, if we now assume that the
probability of getting a fragment of size $a$ out of an urn with $A$
particles is $P(a,A)$, the normalized yield of such a sampling will be: 
\begin{equation}
Y(n,N,z,Z)=P(a,A)P(n,N,z,Z,A)=P(a,A){\frac{{\binom{N}{n}\binom{Z}{z}}}{{%
\binom{A}{a}}}}\ .
\end{equation}%
In particular for the case of sampling $a$ particles composed of $n$
neutrons and $z$ protons ($a=n+z$) from an urn with $A_{1}=Z_{1}+N_{1}$ the
term $P({n,N_{1},z,Z_{1})}$ becomes 
\begin{equation*}
P({n,N_{1},z,Z}_{1},A_{1})={\frac{{\binom{N_{1}}{n}\binom{Z_{1}}{z}}}{{%
\binom{A_{1}}{a}}}}=\frac{N_{1}!}{n!(N_{1}-n)!}\frac{Z_{1}!}{z!(Z_{1}-z)!}{%
\frac{a!(A_{1}-a)!}{A_{1}!}}\ .
\end{equation*}%
And taking the ratio of this probability to the probability of obtaining $n$
neutrons and $z$ protons from a sampling of $a$ nucleons from an urn with $%
A_{2}$ nucleons composed of $N_{2}$ neutrons and $Z_{2}$ protons, the
isoscaling ratio is given exactly by 
\begin{eqnarray}
R_{21} &=&\frac{P(a,A_{2})}{P(a,A_{1})}\frac{P(n,N_{2},z,Z_{2},A_{2})}{%
P(n,N_{1},z,Z_{1},A_{1})}  \label{eq4} \\
&=&\frac{P(a,A_{2})}{P(a,A_{1})}{\frac{{\binom{N_{2}}{n}\binom{Z_{2}}{z}}}{%
\binom{A_{2}}{a}}}{\frac{{\binom{A_{1}}{a}}}{{\binom{N_{1}}{n}\binom{Z_{1}}{z%
}}}}\ ,
\end{eqnarray}%
which can be readily used for calculations taking $P(a,A_{2})/P(a,A_{1})$ as
an overall normalization.

This expression, although close to the usual exponential law~(\ref{eq1}), is
not a straight line in the linear-log plot of $R_{21}$ versus $N$; the fact
that experimental data also deviates from such a linear behavior is
reassuring (see $eg.$~\cite{mockot}). The approximate exponential law~(\ref%
{eq1}) can be obtained from the exact result~(\ref{eq4}) using the binomial
approximation to the hypergeometric distribution (see \textit{eg.}~\cite%
{ross}) which, in this case, depends on the assumption that $x\ll X_{i}$ for 
$x=a,n,z$, $X_{i}=A_{i},N_{i},Z_{i}$ and $i=1,2$. The sampling isoscaling
ratio then becomes 
\begin{equation}
R_{21}\approx \frac{P(a,A_{2})}{P(a,A_{1})}\exp \left\{ n\ln \left( \frac{%
q_{2}}{q_{1}}\right) +z\ln \left( \frac{p_{2}}{p_{1}}\right) \right\} \ ,
\label{eq5}
\end{equation}%
where we have introduced the probabilities of extracting a neutron, $%
q_{i}=N_{i}/A_{i}$, and a proton, $p_{i}=Z_{i}/A_{i}$.

Comparing to equation~(\ref{eq1}) we identify the overall normalization
constant as $C=P(a,A_2)/P(a,A_1)$ and the isoscaling parameters as $%
\alpha(S)=\ln (q_2/q_1)=\ln (N_2A_1/N_1A_2)$, $\beta(S)=\ln (p_2/p_1)=\ln
(Z_2A_1/Z_1A_2)$, in perfect agreement with the percolation results~\cite%
{davila,dorso-1}. Reviewing the procedure leading to equation~(\ref{eq5}),
it is clear that exponential law is a direct result of the sampling.

[For completeness, although not relevant to the nuclear case, we note that a
functionally similar result can be obtained for the case of sampling with
replacement. In this case, fragments with $a=n+z$ will appear with
probability $P(n,N,z,Z,A)= Cp^{z}q^{n}= C p^{z}(1-p)^{a-z}$, where $C$ is
given by the normalization $C ^{-1}=\left( p^{A+1}-q^{A+1}\right)/(p-q)$,
with $p$ and $q$ were defined before. Using this for urns $1$ and $2$ leads
to 
\begin{equation}
R_{21}=\frac{P(a,A_2)}{P(a,A_1)} \left[ \frac{p_2^zq_2^n}{p_1^zq_1^n}\right] %
\left[ \frac{\left( p_2-q_2\right) \left( -q_1^{a+1}+p_1^{a+1}\right) }{%
\left( p_1-q_1\right) \left( -q_2^{a+1}+p_2^{a+1}\right) }\right]=C(a)\frac{%
P(a,A_2)}{P(a,A_1)} \left[ \frac{q_2}{q_1}\right]^n\left[ \frac{ p_2}{p_1}%
\right] ^z  \label{eq6}
\end{equation}
where $C(a)=\left[\left( 2p_2-1\right) \left( -q_1^{a+1}+p_1^{a+1}\right) %
\right]/\left[ \left( 2p_1-1\right) \left( -q_2^{a+1}+p_2^{a+1}\right) %
\right]$. Since $C(a)$ is independent of $n$ and $z$, equation~(\ref{eq6})
depends on these variables in a functionally similar manner as the
isoscaling law~(\ref{eq1}).]

\section{Comparison to experimental isoscaling}

\label{sec2}

The energy dependence of the isoscaling parameters has already been explored
in collisions~\cite {dorso-CMD}; here we compare the 
isoscaling parameter $\alpha $ obtained
from experiments to those from samplings, and study their variation as a
function of beam energy.\bigskip

In view of the previous result, namely the fact that isoscaling can be
expected from the mere act of fragmenting a system, the question to answer
now is: what fraction of the nuclear isoscaling is due to sampling?

In what follows we will show to which extent combinatorial effects (aka
symentropy effects, see \cite{moretto-dorso}), are relevant by comparing the value
of $\alpha $ obtained from experimental calculations with the one resulting
from the simple combinaorial analysis. It is worth realizing at this point
that in the case of the combinatorial analysis we are disregarding all
correlations , in particular those associated with energy terms and then
correspond to the very high Temperature limit.(see for example \cite{dorso-1}%
).We then expect that the contribution of combinatorial terms to be apparent
in collisions at high energies. Here we present a direct comparison of our
findings to two sets of experimental data --not to try to reproduce them
but-- to assess the relative magnitude of the sampling contribution to
isoscaling, and to attempt to draw a baseline from which future experimental
studies will be able to extract the nuclear contribution to this effect.

The data used was obtained by Yennello \textit{et al.}~\cite{yenello} at the
Cyclotron Institute of the Texas A\&M University

\begin{figure}[tbp]
\centerline{ \includegraphics[scale=0.5,angle=0]{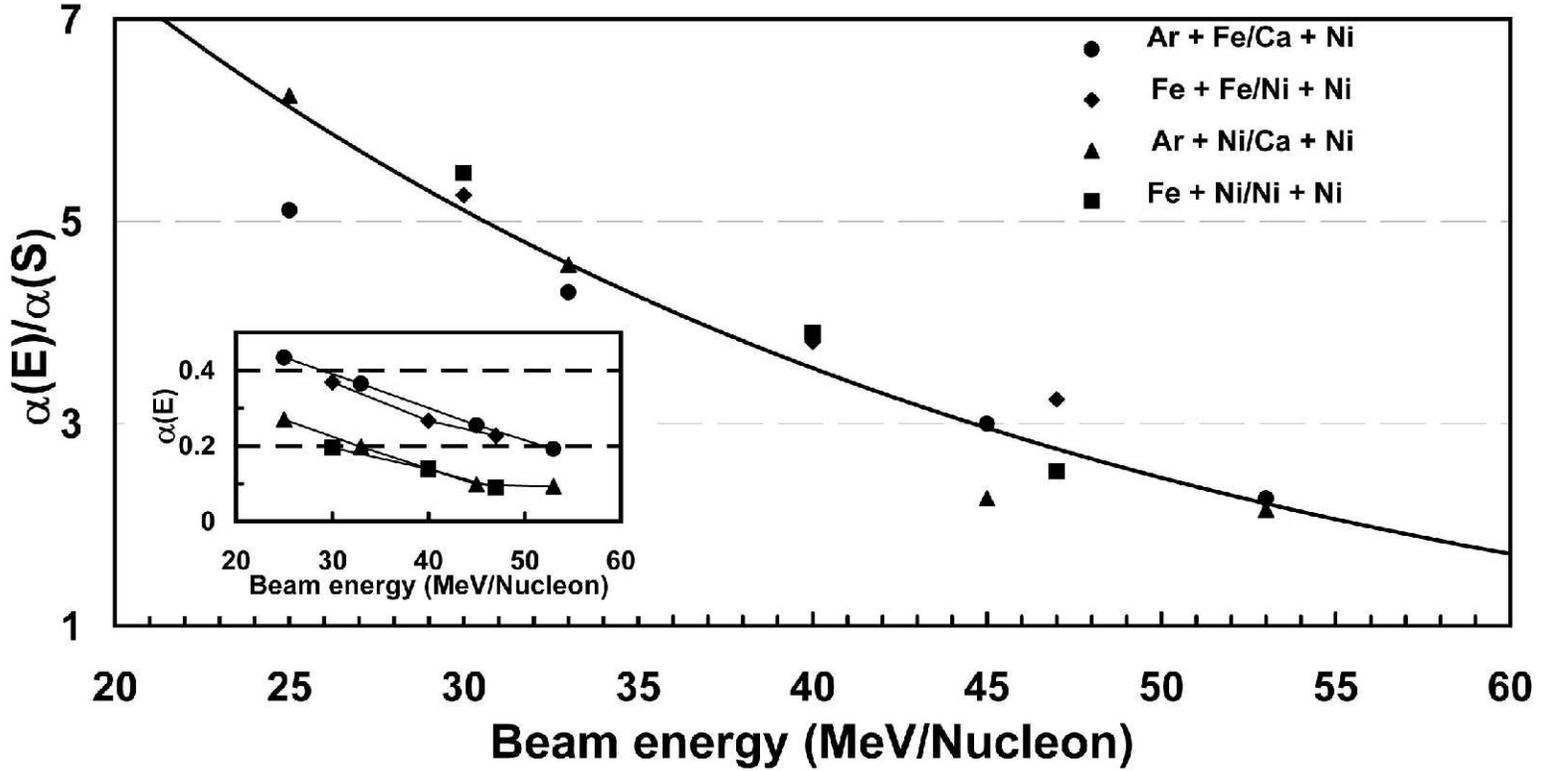}}
\caption{Ratio of experimental $\protect\alpha(E)$ to the sampling $\protect%
\alpha(S)$ as a function of energy.}
\label{fig1}
\end{figure}

The values of $\alpha (E)$ used in this comparison are shown in the inset of
figure~\ref{fig1} and correspond to the ratios of the yields $%
Y({}^{40}Ar+{}^{58}Fe)/Y({}^{40}Ca+{}^{58}Ni)$, $%
Y({}^{58}Fe+{}^{58}Fe)/Y({}^{58}Ni+{}^{58}Ni)$, $%
Y({}^{40}Ar+{}^{58}Ni)/Y({}^{40}Ca+{}^{58}Ni)$ and $%
Y({}^{58}Fe+{}^{58}Ni)/Y({}^{58}Ni+{}^{58}Ni)$ at the energies shown. The
main panel of figure~\ref{fig1} shows the ratio of $\alpha (E)$ to the
corresponding parameter obtained from the sampling, $\alpha (S)=\ln
(N_{2}A_{1}/N_{1}A_{2})$; \textit{cf.} equation~(\ref{eq5}).

It is easy to see that at all energies $\alpha (E)>\alpha (S)$, and that for
large energies $\alpha (E)\rightarrow \alpha (S)$. A second point of
interest is the fact that upon division by $\alpha (S)$, the previously
scattered four curves of $\alpha (E)$ values collapse into a single group..

\section{Conclusions}

The conclusions of this work are quite simple. First we note that isoscaling
can be expected to appear in any system undergoing a \textit{fair} sampling;
nuclear fragmentation --being a type of sampling-- is bound to exhibit this
phenomenon.

Second, the inherent correlations of nuclear systems are expected to have an
effect on the \textit{fairness} of the sampling modifying the yield ratio $%
R_{21}$ and the isoscaling parameters $\alpha$ and $\beta$, as was
demonstrated by a direct comparison to experimental results.

Finally, a word of caution is needed if one attempts to extract information
about the nuclear equation of state from isoscaling. Given that it is now
known that a sampling-related isoscaling is ever present, obtaining
quantities such as $C_{sym}$ from equation~\ref{eq2} is not straightforward.
In principle, the experimental results should contain the isoscaling
produced by sampling, and --in some regime-- both $R_{21}(E)$ and $\alpha(E)$
should tend to $R_{21}(S)$ and $\alpha(S)$.

The indication that $\alpha(E) \rightarrow \alpha(S)$ at high energies
indicates --perhaps-- that at those energies nuclear binding is less
important and the ``\textit{sampling}'' is closer to that of a
non-interacting urn; at lower energies, however, the reaction has longer
interaction times and the phase space available for the \textit{sampling}
becomes a complex function of the energy distancing itself from the fair
sampling case of equations~(\ref{eq4}) and~(\ref{eq5}). This is consistent
with the implications of equation~(\ref{eq2}) which, through its inverse
dependence on the temperature, indicates that in the limit of high energies
the expected contribution from the equation of state to the isoscaling
coefficients should vanish: $\alpha \rightarrow 0$. Likewise, studies of
bond percolation on polychromatic substrates~\cite{dorso-1,moretto-dorso}
(generated through the nuclear lattice model at temperature $T$) have shown
that in the limit of high temperatures only the probabilistic term survives.

\textbf{Acknowledgements} \newline
J.A.L. thanks the University of Colima, and C.O.D. the University of Texas
at El Paso, for their hospitality while part of this work was carried out.

\end{document}